\documentclass[acmsmall,screen]{acmart}

\setcopyright{none}
\renewcommand\footnotetextcopyrightpermission[1]{} 
\usepackage[utf8]{inputenc}
\usepackage[T1]{fontenc}
\usepackage{longtable} 
\usepackage{multirow}
\usepackage{ragged2e}
\usepackage{tabularx}
\usepackage{graphicx}      
\usepackage{geometry}
\usepackage{subcaption}   
\usepackage{array} 
\usepackage{xltabular} 
\usepackage{makecell}
\usepackage{float}
\usepackage{booktabs}
\usepackage{rotating}
\usepackage[normalem]{ulem}

\usepackage{minted} 
\newcommand{\tabincell}[2]{\begin{tabular}{@{}#1@{}}#2\end{tabular}}

\title{Understanding and Detecting Flaky Builds in GitHub Actions}

\author{Wenhao Ge}
\email{whge@stu.suda.edu.cn}
\affiliation{%
  \institution{School of Computer Science and Technology, Soochow University}
  \city{Suzhou}
  \country{China}
}

\author{Chen Zhang}
\email{chenzhang22@stu.suda.edu.cn}
\affiliation{%
  \institution{School of Computer Science and Technology, Soochow University}
  \city{Suzhou}
  \country{China}
}

\begin{abstract}
Continuous Integration (CI) is widely used to provide rapid feedback on code changes; however, CI build outcomes are not always reliable. Builds may fail intermittently due to non-deterministic factors, leading to flaky builds that undermine developers’ trust in CI, waste computational resources, and threaten the validity of CI-related empirical studies. In this paper, we present a large-scale empirical study of flaky builds in GitHub Actions based on rerun data from 1,960 open-source Java projects. Our results show that 3.2\% of builds are rerun, and 67.73\% of these rerun builds exhibit flaky behavior, affecting 1,055 (51.28\%) of the projects. Through an in-depth failure analysis, we identify 15 distinct categories of flaky failures, among which flaky tests, network issues, and dependency resolution issues are the most prevalent. Building on these findings, we propose a machine learning–based approach for detecting flaky failures at the job level. Compared with a state-of-the-art baseline, our approach improves the F1-score by up to 20.3\%.
\end{abstract}

\ccsdesc[500]{Software and its engineering~Software configuration management and version control systems}

\keywords{Continuous Integration, Flaky Build, GitHub Actions, Rerun Build}

\begin{document}

\maketitle

% --- 正文导入 ---
% 请确保你的项目目录下存在 src 文件夹，且包含以下 .tex 文件
% !TeX root = ../main.tex

\section{Introduction}

Continuous Integration (CI) is a cornerstone of modern collaborative software development~\cite{golzadeh2022rise}. CI automatically builds code changes to identify potential bugs before they are integrated into a central repository. CI is widely adopted due to its many benefits, such as detecting bugs earlier and accelerating development cycles~\cite{Hilton2017Trade,vasilescu2015quality}. With the growing importance of CI, a rich ecosystem of CI services has emerged. Among them, GitHub Actions~\cite{github-action.com} is one of the most widely used. Since its introduction in November 2019, GitHub Actions has quickly gained popularity and become the leading CI service, owing to its seamless integration with GitHub and generous free tier~\cite{golzadeh2022rise, rostami2023usage}.

CI provides rapid feedback on build outcomes, which developers rely on to plan their subsequent work. When a build succeeds, developers can confidently proceed with subsequent development tasks. When a build fails, developers may need to pause and diagnose issues potentially introduced by the code changes~\cite{maipradit2023repeated}. Although build outcomes serve as important indicators, they are not always reliable. CI builds may fail intermittently due to non-deterministic factors, such as network instability and flaky tests (i.e., tests that unpredictably pass or fail without any code changes)~\cite{luo2014empirical}.
In such cases, the build failures should not be attributed to the code changes. Many CI services, including GitHub Actions, provide a rerun mechanism for re-executing CI builds~\cite{durieux2020empirical, maipradit2023repeated}.
%Depending on the platform, this feature may be labeled as \textit{restart}, \textit{recheck}, or \textit{repeat}~\cite{durieux2020empirical, maipradit2023repeated}. 
When developers suspect that a build outcome does not align with their expectations, they may rerun the build to determine whether the failure was caused by bugs introduced by the code change or by non-deterministic factors~\cite{aidasso2025illusion}. 

The unreliability of CI builds challenges the assumption that a build failure indicates a bug in the code change and erodes developers' trust in CI. It also leads to wasted effort and resources, as developers must spend additional time investigating failures and rerunning builds to identify their root causes~\cite{maipradit2023repeated}.
Moreover, such unreliability threatens the validity of empirical studies on CI builds~\cite{ghaleb2019studying} and build outcome prediction models~\cite{chen2020buildfast}, which typically assume that build outcomes are reliable. Following prior work~\cite{durieux2020empirical}, we define builds that change their outcomes after reruns without any code changes as \textbf{flaky builds}. Given the negative impacts of flaky builds, it is critical to understand their frequency and root causes, and to develop automated techniques to help CI users detect them.

Given the availability of detailed rerun data in GitHub Actions, we leverage rerun records from 1,960 open-source Java projects. As part of our empirical study, we first investigate the frequency of flaky builds and find that 3.2\% of builds are rerun, among which 67.73\% are flaky builds. We then examine flaky failures and identify 15 failure categories, with flaky tests, network issues, and dependency resolution issues being the most prevalent. Building on these empirical findings, we propose a machine learning–based approach for detecting flaky failures. Compared with a strong state-of-the-art baseline~\cite{olewicki2022towards}, our approach improves the F1-score by up to 20.3\%.

To the best of our knowledge, studies by Durieux et al.~\cite{durieux2020empirical} and Maipradit et al.~\cite{maipradit2023repeated} are the only works that examine the rerun feature in CI. However, they do not specifically target GitHub Actions. Prior work has also explored the detection of flaky failures~\cite{lampel2021life,olewicki2022towards}; however, these studies primarily target industrial CI systems and rely on features that are often unavailable in open-source environments. In summary, this paper makes the following three key contributions.

\begin{itemize}
\item We construct and release two GitHub Actions datasets: (1) a large-scale CI build dataset collected from 1,960 open-source Java projects, and (2) a curated flaky failure dataset derived from 10 projects.
    
\item We identify and systematically characterize flaky failure types in GitHub Actions.
    
\item We propose a machine learning–based approach for detecting flaky failures which significantly improves F1-score.
\end{itemize}

\section{Background}\label{sec:background}

%In this section, we provide an overview of the background on GitHub Actions and flaky builds.

%\textbf{GitHub Actions.} GitHub is the most popular collaborative development platform, hosting more than 90 million public open-source projects\footnote{https://api.github.com/search/repositories?q=is:public+fork:true}. Recognizing the critical role of CI in modern software development, GitHub introduced GitHub Actions in November 2019 as an officially supported and tightly integrated CI service within the platform. GitHub Actions allows developers to automate a variety of tasks, including compilation, testing, static analysis, and deployment. Before the introduction of GitHub Actions, Travis CI was the predominant CI service on GitHub~\cite{hilton2016usage}. However, changes in Travis CI's pricing policy in 2020 imposed stricter limitations on its free tier, which prompted many users to seek alternatives.
%Remarkably, only 18 months after its introduction, GitHub Actions became the dominant CI service on GitHub~\cite{golzadeh2022rise}.
%Many projects have used GitHub Actions to compile, test, analyze and deploy code~\cite{kinsman2021software}.
%Thus, we focus on GitHub Actions in this paper to maximize the impact of our findings.

\textbf{Workflows, Builds, and Jobs.} In GitHub Actions, a \textbf{workflow} is an automated process composed of one or more jobs~\cite{github-action.com}. Workflows are defined using YAML configuration files located in the \texttt{.github/workflows} directory of a repository. GitHub Actions workflows can be triggered by a wide range of developer activities. These triggers, also known as events, include common activities such as code pushes, issue openings, and pull request creation, as well as more specialized triggers like scheduled times or manual invocations. 
%Additionally, workflows can be triggered by external events outside of GitHub through webhooks. 
%For instance, in Fig.~\ref{lst:workflow-example}, according to the \texttt{on} key specified in line 2, the workflow is executed whenever a pull request is created....
In this paper, we use the term \textbf{build} to denote the execution of a workflow, consistent with terminology adopted in prior work~\cite{durieux2020empirical, ghaleb2019studying}.
% \begin{figure}[!t]
%     \centering
%     \includegraphics[width=1\linewidth]{src/images/wf_exam.pdf}
%     \vspace{-10pt}
%     \caption{An Example Workflow Configuration}
%     \label{fig:workflow-example}
%     \vspace{-10pt}
% \end{figure}
A \textbf{job} is a collection of steps designed to perform specific tasks (e.g., compilation, testing, static analysis, or deploying artifacts). Each step within a job can be either a shell script or an action. An action is a reusable extension that can be shared and reused across different workflows and repositories. 
%By encapsulating complex or repetitive code as an action, developers can significantly reduce the effort required to configure and maintain GitHub Actions workflows. 
%As shown in Fig.~\ref{fig:workflow-example}, 
%As of December 2022, there are more than 16,730 actions available on the GitHub Marketplace~\cite{saroar2023developers}, which greatly enhances the capabilities of GitHub Actions. 
Each build may contain one or more jobs, and different jobs within the same build can run in different environments, such as distinct operating systems or runtime versions.

\textbf{Outcomes.} GitHub Actions uses two metadata fields, \texttt{status} and \texttt{conclusion}, to represent the execution progress and the final outcome of a build or job\footnote{https://docs.github.com/en/pull-requests/collaborating-with-pull-requests/collaborating-on-repositories-with-code-quality-features/about-status-checks}. The \texttt{status} field indicates the execution progress, with possible values such as \texttt{queued}, \texttt{pending}, \texttt{waiting}, \texttt{in\_progress}, and \texttt{completed}. A value of \texttt{completed} indicates that the build or job has finished execution, whereas the other values denote that it is awaiting resources or approval, or is actively running. Once the \texttt{status} of a build or job becomes \textsf{completed}, the \texttt{conclusion} field is assigned to indicate the final outcome. A value of \texttt{success} indicates that the build or job completed successfully, whereas \texttt{failure} indicates that errors occurred during execution. The value \texttt{startup\_failure} represents configuration issues that prevent the build from starting. The value \texttt{action\_required} signals that manual intervention is required to continue execution. The value \texttt{neutral} indicates that the build or job completed with a neutral outcome.
The \texttt{conclusion} field may also take values such as \texttt{canceled} or \texttt{skipped}, indicating that the execution was canceled or skipped.

%The \texttt{status} key reflects the ongoing state of a build or job, with possible values including \textsf{queued}, \textsf{in\_progress}, \textsf{requested}, \textsf{waiting}, \textsf{pending}, or \textsf{completed}. 

\begin{figure}[!t]
    \centering
    \centerline{\includegraphics[width=0.9\textwidth]{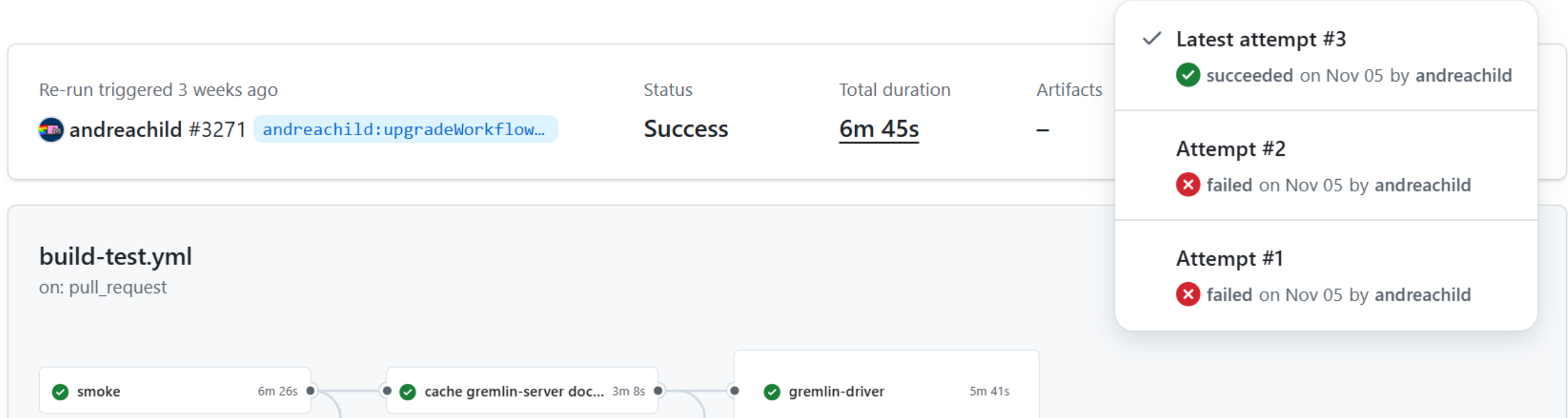}}
    %\vspace{-10pt}
    \vspace{-5pt}
    \caption{An Example of a Rerun Build That Exhibits Flakiness} 
    \label{fig:github-rerun-example}
    \vspace{-20pt}
\end{figure}

\textbf{Rerun Builds and Flaky Builds.} GitHub Actions allows developers to rerun an entire workflow or specific jobs within 30 days after the initial run. If a build is rerun one or more times, we refer to it as a \textbf{rerun build}.
Figure~\ref{fig:github-rerun-example} illustrates an example of a rerun build in GitHub Actions. In this example, the initial execution (Attempt~\#1) failed, the first rerun (Attempt~\#2) also failed, while the second rerun (Attempt~\#3) succeeded.
Different CI services use different terminology to refer to rerun builds. In Travis CI, rerun builds are called restart builds~\cite{durieux2020empirical}, whereas in the OpenStack community, they are known as repeated builds or rechecked builds~\cite{maipradit2023repeated}.

% \begin{figure}[!t]
%     \centering
%     \includegraphics[width=0.8\linewidth]{src/images/flaky-example.pdf}
%     \caption{An Example of Flaky Build}
%     \label{fig:flaky-example}
% \end{figure}

Since a build may contain multiple jobs, and each job performs a specific task with an independent outcome, different jobs within the same build may behave differently. We define a \textbf{flaky job} as a job that exhibits both success and failure across different executions. We define a \textbf{flaky build} as a rerun build that contains at least one flaky job. For the purpose of failure analysis, we further refer to each failure occurrence observed in a flaky job as a \textbf{flaky failure}. In our definition, we exclude outcomes such as \texttt{startup\_failure}, \texttt{action\_required}, \texttt{canceled}, and \texttt{skipped}, as these represent incomplete executions.
Different organizations use domain-specific terminology to describe flakiness. For instance, at Mozilla, flaky jobs are referred to as oranges~\cite{lampel2021life}. At Ubisoft, a flaky build is known as a brown build~\cite{olewicki2022towards}.

%In practice, if a build is rerun enough times, it will eventually exhibit flakiness, as environmental factors such as network errors are inevitable. 
%\textbf{Github Actions VS. Travis CI} Before the introduction of GitHub Actions, Travis CI is the predominant CI service. Additionally, the only work closed to our study is on Travis CI. Thus, we list the differences of GitHub Actions with Travis CI. The configuration file of Travis CI is \texttt{.travis.yml}. A repository can contain multiple workflow files but only one \texttt{.travis.yml}. The events of GitHub Actions are far more strong than Travis CI.

\section{Empirical Study Design}

In this section, we present the design of our empirical study. Since our analysis of flaky builds relies on rerun data from GitHub Actions, we first describe the selection of open-source projects, the collection of build data, and the identification of rerun and flaky builds. We then introduce the metrics and procedures used to quantify the costs of rerun and flaky builds and to identify their root causes. Based on this study design, we address the following two research questions.

\noindent\textit{\textbf{RQ1: Frequency Analysis.} How frequently do flaky builds occur in rerun CI builds?}
Understanding the frequency of flaky builds is essential for assessing the prevalence of flakiness in practice and for motivating the significance of this study.

\noindent\textit{\textbf{RQ2: Failure Analysis.} What failure types lead to flaky builds?}
Flaky builds can arise from diverse underlying causes. By investigating flaky failures, we can better characterize the nature of flakiness and derive actionable insights for mitigating these issues.

\subsection{Data Preparation}

\textbf{Project Selection.} To construct a dataset of flaky builds in GitHub Actions, we used the SEART GitHub Search Engine (seart-ghs)~\cite{dabic2021sampling} to obtain a list of candidate open-source projects hosted on GitHub. We selected seart-ghs because it provides multiple filtering criteria that facilitate the sampling of relevant projects, and its dataset is continuously updated. As of December 2, 2023, seart-ghs indexed 1,543,046 projects.

We first excluded forks, as their data may duplicate that of their upstream projects, resulting in 1,494,717 projects. Since our study specifically targets Java, we next filtered the results to include projects that primarily use the Java programming language, yielding 100,426 projects. To ensure the quality of the projects and exclude experimental ones, we removed those with fewer than 150 stars and 250 commits, narrowing the list to 6,452 projects. 
Because GitHub Actions retains a project's build history for only three months by default, we retained only those projects with at least one commit in the last three months to ensure they were still active and that their build data could be collected, which resulted in 2,746 projects. 
For each of the 2,746 open-source Java projects, we cloned the repositories and verified the presence of the \textsf{.github/workflows} directory, a prerequisite for running GitHub Actions workflows. Finally, we obtained a candidate set of 2,042 projects.

\textbf{Build Data Collection.} To collect the CI build histories of the 2,042 projects, we developed a crawler that interacts with the GitHub REST API. For each build, the collected data consist of three types of files: a build-level JSON file, a job-level JSON file that summarizes all jobs in the build, and several log files. The build-level JSON file contains metadata for the entire build, including the build ID, status, and duration. The job-level JSON file records metadata for all jobs within the build. The log files record the console output generated during job execution, including error messages and runtime information. Each job has one corresponding log file. We ran our crawler to continuously collect build data from September 3, 2023, to August 11, 2024. Of the 2,042 projects, we were unable to collect build data for 82. These repositories may have been deleted, renamed, or had their CI build histories cleared. Consequently, our final dataset consists of 1,960 repositories, comprising 4,861,768 builds and 15,382,267 jobs.

\textbf{Rerun Build and Flaky Build Identification.} GitHub Actions records rerun information directly in the build metadata. In each build-level JSON file, the field \texttt{run\_attempt} specifies the attempt number of the build. A value of \texttt{run\_attempt} = 1 indicates that the build was executed only once, without any reruns. Conversely, a value greater than 1 indicates that the build was rerun additional times after its initial execution. For example, a value of 5 indicates that the build was rerun four additional times after its initial run. We treat all builds with a \texttt{run\_attempt} value greater than 1 as rerun builds. For each rerun build, we collected all metadata and log files associated with every attempt to reconstruct the complete rerun sequence. To identify flaky builds, we followed our definitions by examining the outcomes of the jobs within each rerun sequence. A job is classified as a flaky job if it exhibits both success and failure across its attempts. A build is classified as a flaky build if it contains at least one flaky job.

%By grouping all executions of the same build (i.e., builds sharing the same ID but having different \textsf{run\_attempt} values), we can reconstruct the complete rerun history for each build.

%Unlike Travis CI, which requires periodic crawling build data (e.g., once a day) and comparison of timestamps to identify restarted builds~\cite{durieux2020empirical}, GitHub Actions provides the \texttt{run\_attempt} field in its metadata. 

\subsection{Cost Measurement}

While rerun builds play a central role in detecting flaky builds, they also introduce additional costs that delay development progress and waste computational resources. Understanding these costs is essential for assessing their impact on the CI process and developer productivity. Accordingly, we use two metrics to measure these costs.
%\TODO{To characterize these costs, we introduce two metrics: feedback delay, which captures the additional time developers must wait before receiving stable CI feedback, and resource overhead, which quantifies the extra CI resources consumed by rerunning builds.}

\textbf{Waiting Time.} When a build is triggered, developers must either wait for CI feedback or switch to other tasks. In either case, their workflow is interrupted, resulting in reduced productivity. Because rerunning builds involves multiple executions, the accumulated waiting time further delays feedback delivery. We use the \textit{waiting time} metric to quantify this feedback delay. This metric is computed as the difference between the \texttt{start\_time} of the initial run and the \texttt{end\_time} of the final rerun.

\textbf{Computational Time.} In GitHub Actions, builds are executed on virtual machines hosted by GitHub. These virtual machines provide environments with the necessary operating systems, software, and dependencies to run CI builds. Computational resources for CI are both expensive and scarce~\cite{maipradit2023repeated}. Each build in GitHub Actions may include multiple jobs, with each job running on its own virtual machine. Thus, resource consumption occurs at the job level. We use the metric \textit{computational time} to quantify the resource cost. Computational time is measured as the compute time of all jobs in a rerun sequence, computed by subtracting the \textsf{start\_time} from the \textsf{end\_time} of each job.

\subsection{Failure Category Labeling}

Each job in GitHub Actions generates a log file that records the console output produced during execution. When a job fails, a failure message is reported at the end of the log, describing what went wrong and reflecting how flaky behavior manifests at the surface level. These failure messages serve as the foundation of our analysis. However, raw failure messages are highly diverse and inconsistent across projects, making direct large-scale analysis infeasible. Therefore, we designed a semi-automated failure symptom extraction pipeline to extract, generalize, and consolidate failure-related information. We began by randomly sampling ten failed jobs and manually inspecting their logs to understand the structure and formatting of failure messages. We found that failure messages often contain project-specific variable tokens, such as file paths, package names, dependency versions, and timestamps. For example, the messages “\textit{Error: failed to download starwhale/starwhale}” and “\textit{Error: failed to download helm/helm}” refer to the same underlying failure. 
To enable large-scale automatic matching, we normalized variable tokens by replacing them with wildcards while retaining the invariant failure-indicative components, thereby forming generalized failure patterns. In this case, both messages were abstracted to the regular expression "\textit{Error: failed to download .*?}". This abstraction makes failure patterns reusable and applicable across logs from different repositories.
We applied the derived failure patterns to the full dataset of flaky jobs to automatically match similar messages. To capture failures not covered by the initial patterns, we randomly examined ten unmatched logs, identified previously unseen flaky failure messages, and manually created new patterns. This process was iterated until no additional meaningful patterns could be generated.

Although failure patterns allow us to match similar log messages automatically, the same underlying failure may still appear in inconsistent forms~\cite{ghaleb2019studying}. For example, a network connectivity issue may be logged as “\textit{Error: Remote host terminated the handshake}”, or “\textit{Error: Failed to fetch .*? Network is unreachable}”. While these messages differ syntactically, they describe the same failure phenomenon. To avoid fragmentation, we grouped related failure patterns into higher-level failure categories. A failure category captures a recurring type of failure that manifests at the surface level and may be represented by multiple failure patterns.

\subsection{Flaky Failure Analysis}

Analyzing root causes is essential for understanding why flaky failures occur beyond their surface manifestations. Given the large number of flaky failure cases in our dataset, manually analyzing all instances is infeasible. We therefore adopted an iterative sampling strategy~\cite{zhang2019large}. In each iteration, we randomly sampled five flaky failure cases from each failure category and manually inspected their code changes and execution logs. When necessary, we additionally consulted related issue reports or documentation to identify previously unseen root causes or mitigation patterns. We repeated this process across multiple iterations and terminated the analysis when two consecutive iterations yielded no new root causes or mitigation patterns, indicating that thematic saturation had been reached.

\section{Empirical Study Results}\label{sec:result}

%In this section, we present the empirical results for RQ1 and RQ2.

\subsection{Frequency Analysis (RQ1)}

\textit{\textbf{Observation 1:} Of the 4,879,460 builds, 155,488 (3.2\%) were rerun.} This proportion was nearly twice the 1.7\% reported by Durieux et al.~\cite{durieux2020empirical}. This discrepancy can be attributed to two primary factors. First, Travis CI does not explicitly record rerun information in its build metadata. As a result, identifying restarted builds in Travis CI requires periodic crawling of build data, which may result in missing builds that are restarted too quickly. Second, GitHub Actions introduces an approval-based execution control mechanism to enhance security and prevent resource abuse. For example, builds triggered by first-time contributors are not allowed to execute automatically and are instead marked with the \textsf{action\_required} outcome. These builds must be manually approved by a maintainer before execution can proceed, after which they are automatically rerun. 

To further examine this second factor, we analyzed the number of reruns per build. Among the 155,488 rerun builds, 142,463 (91.6\%) were rerun exactly once, while 12,945 (8.3\%) were rerun two or more times. A more detailed investigation reveals that 102,878 of the 142,463 single-rerun builds initially had an \textsf{action\_required} outcome. While this mechanism improves security and limits resource consumption, it also introduces additional delays in CI feedback. Specifically, we measured the time elapsed between build triggering and approval for these 102,878 builds and find a total waiting time of 84,529 days (231.59 years), with average and median delays of 20 hours and 1.5 hours, respectively. This observation highlights the need for automated techniques to assist maintainers in deciding whether such builds should be approved promptly.
Because these 102,878 rerun builds are caused by the approval mechanism rather than developer intent, we excluded them from subsequent analyzes. After this exclusion, 52,610 rerun builds remained, affecting 1,186 (60.5\%) of the 1,960 projects, each of which had at least one rerun build. These results indicate that rerunning builds is a widespread practice and affects the majority of projects, reflecting developers' concerns about the reliability of CI build  outcomes. 

\textit{\textbf{Observation 2:} Of the 52,610 rerun builds, 35,634 (67.73\%) exhibited flaky behavior (i.e., flaky builds), affecting 1,005 (51.28\%) of the 1,960 projects.} This result indicates that more than two-thirds of rerun builds changed their outcomes across multiple executions. Moreover, more than half of the projects had at least one flaky build. These results suggest that flakiness is a common challenge in CI, highlighting the importance of studying flaky builds.

While rerunning builds is a common approach for exposing flaky behavior in CI, its effectiveness has not been systematically evaluated. To bridge this gap, we analyze the 35,634 identified flaky builds to determine how many reruns are typically required before an outcome change is observed. Our results show that 10,406 flaky builds required more than one rerun before exhibiting an outcome change, and that extreme cases required up to 20 reruns.

\textit{\textbf{Observation 3:} Waiting for the final outcomes of rerun builds accumulated approximately 339.2 years of waiting time, while rerun executions consumed more than 31.6 years of computational time.} Rerun builds introduce substantial costs to CI in terms of both waiting time and computational time. By aggregating the waiting time of all 52,610 rerun builds, we observed a total waiting time of approximately 2.97 million hours, corresponding to 123,792 days (339.2 years), with average and median waiting times of 19 hours and 2 hours, respectively. In contrast, non-rerun builds exhibited substantially shorter waiting times, with average and median waiting times of approximately 0.3 hours and 2 minutes, respectively. Rerun builds incurred 63-times longer average waiting times and 60-times longer median waiting times than non-rerun builds, and the differences were statistically significant ($p < 0.0001$).

Beyond waiting time, rerunning builds also leads to considerable computational resource consumption. Summing the execution time of all jobs involved in rerun builds, we estimated that they consumed over 31.6 years of cumulative computational time, with average and median computational times of 5 hours and 1.7 hours per build, respectively. In contrast, builds without reruns consumed significantly fewer computational resources. The average and median computational times for non-rerun builds were approximately 0.3 hours and 2 minutes, respectively. Overall, rerun builds consumed 16.7-times longer average and 51-times longer median computational times than non-rerun builds, and the differences were statistically significant ($p < 0.0001$).

Overall, our findings demonstrate that flaky builds are prevalent in CI. More than half of the projects in our dataset were affected by at least one flaky build. Notably, this proportion is likely an underestimation. Our analysis relies on a limited observation window of GitHub Actions build histories, as we were only able to collect build data over several months rather than the complete lifetime of each project. Moreover, our analysis reveals that rerunning builds is not always an effective or efficient strategy. In many cases, flaky behavior does not manifest through reruns or requires multiple reruns. Furthermore, rerun builds incur substantial waiting and computational times, significantly delaying CI feedback and consuming valuable CI resources. These findings highlight the need for a deep understanding of flaky builds and automated techniques to identify them.

% 没有rerun的构建的运行时间统计
% 总的时间 3983693842.0
% 平均时间 862.0
% 中位数时间 50.0

\subsection{Failure Analysis (RQ2)}

In this RQ, we investigate the failure types that lead to flaky builds. Table~\ref{tab:failure-category} provides an overview of the identified failure types. The first column lists the failure categories extracted from build logs, with the corresponding proportions reported in parentheses. The second column presents the associated root causes, while the third column summarizes the mitigation patterns commonly used to address these failures. For the second and third columns, the number of observed cases is reported in parentheses. Our results show that flaky tests constitute the most frequent failure category, accounting for 64.99\% of flaky failures. However, flaky tests have been extensively studied in prior work~\cite{luo2014empirical,pinto2020vocabulary,alshammari2021flakeflagger,fatima2022flakify,gruber2023practical}. Therefore, we focus our analysis on other sources of flaky failures. In total, we identified 15 distinct failure categories. Due to page limitations, Table~\ref{tab:failure-category} reports only the top ten failure categories after excluding flaky tests. The complete list of failure categories is available on our website at \url{https://flaky-build.github.io/}.

\textbf{Network Issue.} This category captures flaky failures caused by temporary network instability during CI job execution and accounts for 15.8\% of the flaky failures. To better understand this category, we manually analyzed 15 cases to identify their underlying root causes and corresponding mitigation patterns. The root causes include request timeouts and connection resets, followed by resource download interruptions, Transport Layer Security (TLS) handshake failures, and connection refused errors. In all observed cases, rerunning the job mitigates the failure by allowing a fresh execution under different network conditions. 

\textbf{Dependency Resolution Issue.} This category captures failures caused by problems in retrieving required dependencies and accounts for 6.32\% of the failure. We manually analyzed 20 cases and identified two root causes. The first root cause is transient network issues that block the download of dependencies from remote repositories. Here, we do not further distinguish them as a separate subtype and instead categorize them under network issues. Such failures are typically mitigated by rerunning the build, which allows the dependency resolution process to be retried under different network conditions.  The second root cause occurs when required dependencies are missing from remote repositories at the time of job execution. Since these repositories are often managed by project maintainers, such failures can be resolved by uploading the missing artifacts to the corresponding remote repositories. This indicates that the failures were caused by incomplete synchronization between the local project and its remote artifact repositories.

\textbf{External Environment Inconsistency.}  This failure category captures flaky failures caused by external environment settings that are not version controlled by the version control system, accounting for 4.7\% of the flaky failures. We manually analyzed 35 cases and identified five root causes.
The most common root cause is workflow policy violations, where jobs fail because policies are not satisfied, such as missing required labels, unapproved pull requests, or pull requests remaining in a draft state. These failures can be mitigated by adding required labels, approving the pull request, or marking it as ready.
The second root cause is authentication failure, which occurs when access tokens expire or users are not included in the allowlist. Such failures can be resolved by refreshing authentication tokens or adding the relevant users to the allowlist.
The third root cause is artifact conflicts, where stale or incompatible artifacts from previous executions interfere with the current build. These failures can be fixed by deleting the conflicting artifacts.
The fourth root cause is intermittent tool failures, in which external tools behave inconsistently across executions. Such failures are typically mitigated by rerunning the build.
The final root cause is upstream repository issues, where temporary inconsistencies or changes in upstream repositories affect the availability or behavior of external resources required during execution. These failures can be mitigated by fixing the affected upstream repositories.

%\textbf{Configuration Inconsistencies.} This failure category refers to build failures caused by temporary inconsistencies or unavailability of external services or upstream repositories that CI builds depend on, and accounts for 2.6\% of the failure cases. These failures arise when upstream providers (e.g., package mirrors, artifact repositories, or signing services) are in an intermediate or inconsistent state, such as during synchronization, replication, or partial updates. As a result, builds may fail intermittently even though neither the source code nor the build configuration has changed. Such failures are typically resolved by rerunning the build after the upstream state stabilizes.

%\textbf{External Status/Data Inconsistencies. } This failure category refers to errors caused by discrepancies between the expected and actual status of external services or data and accounts for 2.1\% of the failure cases.

\textbf{API Service Unavailable.} This failure category refers to build failures caused by exceeding API usage quotas imposed by external services and accounts for 1.16\% of the failure cases. Such failures typically occur when CI jobs issue a large number of API requests within a short time window, triggering service-side rate limiting mechanisms. In practice, these failures are mitigated by waiting for quota recovery before rerunning the build or by switching to alternative authenticated tokens or credentials.

\textbf{Concurrency Issue.} This failure category refers to failures caused by unintended interactions among concurrently executed tasks or threads during CI job execution, accounting for 0.88\% of the observed failure cases. Through a manual inspection of 15 cases, we identified two root causes: lock contention and concurrent collection modification. These failures occur because shared resources are accessed without proper synchronization. In these cases, rerunning the job mitigates the failure by allowing a different execution timing under new conditions.

\textbf{Compilation Error. } This failure category refers to flaky failures that occur during the compilation stage of CI jobs and accounts for 0.82\% of the failures. We manually analyzed 25 cases and identified seven root causes. The most common causes are network issues, which interrupt dependency retrieval during compilation, and corrupted caches, where invalid cached artifacts are restored and interfere with the compilation process. We also observed failures caused by upstream repository issues, in which changes or temporary inconsistencies in upstream repositories affect compilation, as well as stale build caches, where residual artifacts from previous executions pollute the build environment. Additional root causes include runner incompatibility, where the runner does not support required compilation features, intermittent tool failures, and missing dependencies that are unavailable at build time. These failures are typically mitigated by rerunning the job, refreshing or invalidating caches, updating the runner, or uploading the missing artifacts.

\textbf{Execution Crash.} This failure category refers to process or execution environment failing to start or terminating abruptly without a clean exit and account for 0.78\% of the failure cases. Based on a manual analysis of 15 instances, we identified two primary root causes: out of memory(OOM) and unstable runner environments. Our investigation reveals that environmental instability is primarily attributed to poor compatibility between legacy macOS runners and container virtualization tools. Specifically, these older runner occasionally crash during the initialization process. As for OOM, it is triggered by random memory spikes when concurrent tasks overlap. Rerunning works because it changes the execution order, keeping peak memory usage within limits.

\begin{table}[!t]
\centering
\small
\caption{Failure Categories, Root Causes, and Mitigation Patterns of Flaky Builds}
\label{tab:failure-category}
\begin{tabular}{lll}
\toprule
\textbf{Failure Category} & \textbf{Root Cause} & \textbf{Mitigation Pattern} \\
\midrule
\multirow{5}{*}{Network Issue (15.8\%)} & Request Timeout (6) & Rerun (6) \\
\cline{2-3} 
& Connection Reset (3) & Rerun (3) \\
\cline{2-3} 
& Resource Download Interruption (3) & Rerun (3) \\
\cline{2-3} 
& TLS Handshake Failure (2) & Rerun (2) \\
\cline{2-3} 
& Connection Refuse (1) & Rerun (1) \\
\hline

\multirow{2}{*}{\tabincell{l}{Dependency Resolution\\ Issue (6.32\%)}} & Network Issue (9) & Rerun (9) \\
\cline{2-3} 
& Missing Dependency (6) & Upload Artifact (6) \\
\hline

\multirow{8}{*}{\tabincell{l}{External Environment\\ Inconsistency (4.7\%)}} & \multirow{3}{*}{Workflow Policy Violation (16)} & Add Label (12) \\
& & Approve Pull Request (2) \\
& & Mark Pull Request as Ready (2) \\

\cline{2-3}
& \multirow{2}{*}{Authentication Failure (12)} & Refresh Token (8) \\
&  & Add to Allowlist (4) \\

\cline{2-3}
& Artifact Conflict (4) & Delete Artifact (4) \\
\cline{2-3}
& Tool Intermittent Failure (2) & Rerun (2) \\
\cline{2-3}
% & \multirow{2}{*}{Tool Bug (2)} & Rerun* (0) \\
& Upstream Repository Issue (1) & Fix Upstream Issue (1) \\
\hline
\multirow[t]{2}{*}{API Service Unavailable (1.16\%) } & API Rate Limit (15) & Rerun (15) \\
\hline

\multirow{2}{*}{Concurrency Issue (0.88\%)} & Lock Contention (9) & Rerun (9) \\
& Concurrent Collection Modification (6) & Rerun (6) \\
\hline

\multirow{7}{*}{Compilation Error (0.82\%) } & Network Issue (7) & Rerun (7) \\\cline{2-3}

& Corrupted Cache (6) & Refresh Cache (6) \\
\cline{2-3} 
& Upstream Repository Issue (4) & Fix Upstream Issue  (4) \\\cline{2-3}
& Dirty Cache (3) & Refresh Cache (3) \\
\cline{2-3} 
& Runner Incompatibility (2) & Update Runner (2) \\
\cline{2-3} 
& Tool Intermittent Failure (2) & Rerun (2) \\
\cline{2-3} 
& Missing Dependency (1) & Upload Artifact  (1) \\
\hline

\multirow{2}{*}{Execution Crash (0.78\%) } & Out Of Memory (8) & Rerun (8) \\
\cline{2-3} 
& Unstable Runner Environment (7) & Rerun (7) \\
\hline

\multirow{5}{*}{\tabincell{l}{File System Interaction\\ Error (0.71\%)}} & Network Issue (10) & Rerun (10) \\
\cline{2-3} 
& API Rate Limit (3) & Rerun (3) \\
\cline{2-3} 
& External Resource Inconsistency (3) & Update External Resource (3) \\
\cline{2-3} 
& Unstable Cache Key (2) & Rerun (2) \\
\cline{2-3} 
& Authentication Failure (2) & Update Permission (2) \\
\hline

\multirow{5}{*}{Static Analysis Error (0.66\%)} & Upstream Repository Issue (8) & Fix Upstream Issue (8) \\\cline{2-3} 
& Network Issue (6) & Rerun (6) \\
\cline{2-3} 
& External Resource Inconsistency (3) & Update External Resource (3) \\
\cline{2-3} 
& Stale Cache (2) & Refresh Cache (2) \\
\cline{2-3} 
& API Rate Limit (1) & Rerun (1) \\
\hline

\multirow{2}{*}{Memory Limit (0.5\%)} & Heap Memory Exhaustion (11) & Rerun (11) \\
\cline{2-3} 
& Disk Space Exhaustion (4) & Clean Disk Space (4) \\

%\multirow[t]{3}{*}{External Environment Inconsistency (4.7\%)} &

\bottomrule
\end{tabular}
\end{table}

\textbf{File System Interaction Error.} This failure category captures flaky failures arising from interactions with the local file system during CI job execution and accounts for 0.71\% of the observed flaky failures. We manually analyzed 20 cases and identified five root causes. The most common root cause is network issues, where file downloads fail transiently while the job continues execution, leading to inconsistent file system states across reruns. We also observed failures caused by API rate limits, which similarly interrupt file downloads due to request throttling by external services. Other root causes include resource inconsistencies, where externally dependent resources are in an inconsistent or unavailable state and can be resolved by updating the corresponding external resources, and unstable cache keys, which result in inconsistent cache states across executions and are typically mitigated by rerunning the job. Finally, some failures stem from authentication failures, where insufficient permissions prevent access to file resources and can be resolved by updating or refreshing permissions.

\textbf{Static Analysis Error.} This failure category refers to flaky failures observed during the execution of static analysis tools and accounts for 0.66\% of the failures. We manually analyzed 20 cases and identified five root causes.
The most common root cause is upstream repository issues, where failures are introduced by changes or inconsistencies in upstream repositories. Such failures are typically mitigated by fixing the issues in upstream repositories. We also observed network issues, which arise when static analysis tools fetch remote resources or communicate with external services; these transient failures are commonly mitigated by rerunning the build. Other root causes include external resource inconsistencies, where external resources vary across executions and can be resolved by updating the corresponding dependent resources, and stale caches, in which outdated cached artifacts interfere with analysis results and can be addressed by refreshing the cache. Finally, some failures are caused by API rate limits imposed by external services, which are typically mitigated by rerunning the job.

\textbf{Memory Limit.} This category captures failures that occur when a CI job consumes more memory resources than are available in the execution environment, accounting for 0.5\% of the failures. Through manual inspection of 15 cases, we identified two root causes. The first root cause is heap memory exhaustion, which occurs when the runtime heap used for object allocation during build execution exceeds its configured limit. These failures are often triggered by concurrent processes or threads competing for memory resources. In such cases, rerunning the job can mitigate the failure by altering execution order or memory allocation timing. The second root cause is disk space exhaustion, which occurs when the CI runner runs out of available storage during execution. 
This issue is particularly prevalent on self-hosted runners with limited disk capacity or insufficient cleanup mechanisms. 
Such failures can be addressed by cleaning obsolete files or provisioning additional disk space for the CI environment.

Our results show that rerunning jobs is the most common mitigation pattern for flaky failures; however, reruns often serve only as a temporary workaround rather than resolving the underlying root causes. We also observe that the same root cause can manifest as different failure symptoms, making diagnosis difficult when relying solely on surface-level error messages. These findings highlight the need for automated fixing tools. Moreover, flaky failures frequently originate from factors beyond the current repository, such as upstream repositories or external services, suggesting that code changes should be coordinated with changes in the surrounding CI ecosystem.

\scriptsize 
\setlength{\tabcolsep}{3pt} 
\renewcommand{\arraystretch}{1.2} 

\large
\section{Flake Detector}\label{sec:tool}

According to the findings of RQ1 and RQ2, flaky builds are common in CI, and relying solely on rerunning builds to address them is both costly and ineffective. To assist developers in efficiently identifying which failed jobs are caused by flaky failures, we propose a machine learning–based approach and develop a corresponding automated detection tool, \textsc{FlakeDetector}, to investigate the following research question.

\noindent\textit{\textbf{RQ3: Effectiveness Analysis.} To what extent can we predict whether a failed job is caused by a flaky failure?} Since a build may consist of multiple jobs, not all jobs within a flaky build are necessarily flaky. Therefore, we perform flaky failure at the job level.
%\underline{Motivation.} The preceding analyses (RQ1 and RQ2) have quantified the non-trivial costs and characterized the complex failure mechanisms of flaky builds. Although the build has concluded, accurately determining whether a specific failure stems from a \textbf{flaky issue} (which might be safely retried or ignored) or a \textbf{deterministic bug} (requiring immediate developer attention) remains a critical bottleneck in the CI/CD pipeline. Developers often waste significant time manually analyzing and attempting to reproduce transient, non-deterministic failures. Therefore, our aim is to leverage a comprehensive set of features—including static code characteristics, build metadata, and historical execution records—to train a machine learning model for \textbf{fast and accurate post-build failure classification}. This high-fidelity diagnostic capability is expected to drastically reduce the time and resources spent on unnecessary debugging efforts, thereby optimizing overall development efficiency and cost.
%\underline{Results.}
% 这里要说明模型是失败构建后预测当前的构建是否是flaky的，注意在上个章节分析完之后说明通过具体分析，flaky是复杂的，仅仅通过日志输出判断可能...，所以使用模型来预测
\subsection{Data Collection}

To construct and evaluate \textsc{FlakeDetector}, we require a labeled dataset of failed jobs, where each failure is classified as either a flaky failure or a non-flaky failure. Accordingly, we retained only jobs whose initial execution resulted in failure. We first examined the rerun history of these failed jobs. If a failed job was rerun by developers and subsequently succeeded, we directly labeled the corresponding failure as a flaky failure, following standard practice in prior work~\cite{lampel2021life,olewicki2022towards}. However, many failed jobs were rerun only a limited number of times or were not rerun at all. Treating all such failures as non-flaky would likely introduce a substantial number of false negatives, as flaky failures with low failure rates may not manifest a successful outcome within a small number of reruns.

To mitigate this issue, we developed CI-Rerunner, an automated tool that systematically reproduces and reruns failed jobs under the same code version. For each failed job that had not been observed to succeed after developer-triggered reruns, CI-Rerunner reran the job up to 10 times, terminating immediately once a successful execution was observed. If any rerun succeeded, the failure was labeled as a flaky failure; otherwise, it was labeled as a non-flaky failure.

Our preliminary analysis showed that 99\% of flaky jobs succeeded within five reruns, consistent with observations that most flaky failures exhibit relatively high recovery probabilities once rerun. To capture a small number of harder-to-detect cases while keeping computational costs manageable, we conservatively increased the rerun limit to ten. This automated multi-rerun strategy identified 83 additional flaky jobs that would have been missed by relying solely on developer-triggered reruns.

Nevertheless, we acknowledge that even ten automated reruns may still fail to expose extremely low-probability flaky behaviors. To assess the extent of such remaining false negatives, we conducted an additional manual analysis phase. Specifically, after applying CI-Rerunner, we randomly sampled 50 jobs that consistently failed across all ten reruns and manually inspected their logs and execution contexts. This analysis revealed 4 additional flaky jobs that had not succeeded during automated reruns, indicating that a small number of flaky failures can remain undetected even after extensive rerunning.

These results confirm that manual auditing is more accurate for identifying flaky failures, but also highlight its impracticality at scale due to its high cost and limited throughput. In contrast, our automated multi-rerun approach significantly improves labeling quality over standard practices while remaining scalable for large-scale studies. We therefore view CI-Rerunner as a practical compromise between labeling accuracy and feasibility: although it does not fully eliminate false negatives, it substantially reduces them and provides sufficiently reliable ground truth for training and evaluating FlakeDetector.

% \begin{figure}[H]
%     \centering
%     \includegraphics[width=1\linewidth]{src/images/ci-reproducer.pdf}
%     \caption{CI-Reproducer}
%     \label{fig:ci-reproducer}
% \end{figure}

\subsection{Feature Engineering}
\label{sec:semantic-log-analysis}

\textbf{Preprocessing and Vectorization.} Raw build logs are full of standard process outputs that appear in every build. These repeated lines are noise. They hide the real error information and confuse the model. To fix this, we filter the logs using regular expressions. We specifically keep lines that match critical error patterns, such as lines starting with error markers (e.g., "E:", "FATAL") or structured log entries indicating failure (e.g., level=fatal). This step removes irrelevant info and helps us find the real failure causes.

Next, we normalize the text. We replace changing parts—such as timestamps, memory addresses, and URLs—with simple placeholders. This is important because these details change in every run. Removing them helps the model learn the general error structure instead of specific details from one build. Finally, we use \texttt{SentenceTransformer all-MiniLM-L6-v2}~\cite{reimers2019sentence} to turn the cleaned text into vectors. This captures the meaning of the errors, allowing us to find similar patterns even if the wording is slightly different. We store these vectors in a FAISS database~\cite{johnson2019billion} for quick search.

\textbf{Feature Representation and Scoring.} To prevent data leakage and simulate a real-world scenario where future data is unavailable, we construct the vector database exclusively using the training dataset. For each target build, we retrieve the top-$K$ most similar historical logs, since similar log patterns typically indicate the same type of failure. $K$ is a hyperparameter optimized in Section \ref{sec:validation}. 

Then, we construct a feature vector $X$ by concatenating the similarity scores ($S_i$) and ground-truth labels ($L_i$) of these neighbors. Given the low dimensionality of this feature vector, we employ a Logistic Regression model~\cite{cox1958regression} rather than a complex neural network. This lightweight choice avoids overfitting and naturally generates a probability between 0 and 1, which serves as an ideal confidence score to quantify the likelihood of flakiness.

\subsection{Structured Feature Analysis}
\label{sec:structured-feature-analysis}

While log analysis captures runtime errors, it may miss the root causes rooted in code changes or environmental contexts. To complement the semantic log features, we developed a specialized static analysis tool, named \texttt{CI-Miner}, to automatically extract structured features from the CI/CD pipeline.

\textbf{Feature Extraction.} We constructed a comprehensive feature set by combining standard metrics from prior studies with new features identified during our root cause analysis. As detailed in Table~\ref{tab:full_features_integrated}, \texttt{CI-Miner} extracts these metrics by parsing git metadata, analyzing abstract syntax trees (AST), and processing CI logs. We organized these features into three dimensions:

\textit{Developer Dimension.} This dimension quantifies human factors using metrics such as committer experience and \textit{Bayesian trust scores} (e.g., \texttt{gh\_committer\_trust\_recent}). We include these features because developers with less experience are often unfamiliar with project-specific testing norms. Similarly, our trust score reflects a developer's historical reliability; a low score indicates a history of introducing regressions, which naturally increases the risk of submitting flaky code.

\textit{Code Change Dimension.} This dimension measures the size and structure of the change. We look at complexity metrics like code churn and entropy, because large or scattered changes are hard to test alone and often cause side effects. Furthermore, we look deeper than just counting lines by using AST analysis (e.g., checking if method signatures changed). This helps us see the real logic of the change. Modifying core logic is much riskier and more likely to break hidden connections between modules than just updating documentation.

\textit{Project \& Context Dimension.} This dimension looks at the project's history and environment to check for outside instability. We track stability (like recent failure rates) and project age, because older projects often have accumulated problems that cause crashes. We also watch for environment changes (like \textit{Dockerfile} updates) and system load (like concurrent jobs). Environment changes can cause compatibility conflicts, while high load stresses system resources, leading to timeouts and flakiness.

\begin{tiny}
\renewcommand{\arraystretch}{1.3}
\begin{xltabular}{\textwidth}{lX @{\hspace{0.4cm}} lX}
    % --- 表头设置 ---
    \caption{Features across Three Dimensions} \label{tab:full_features_integrated} \\
    \toprule
    \textbf{Feature} & \textbf{Description} & \textbf{Feature} & \textbf{Description} \\
    \midrule
    \endfirsthead
    
    % --- 续页表头 ---
    \multicolumn{4}{c}{Table \ref{tab:full_features_integrated} (Continued)} \\
    \toprule
    \textbf{Feature} & \textbf{Description} & \textbf{Feature} & \textbf{Description} \\
    \midrule
    \endhead
    
    % --- 续页表尾 ---
    \midrule
    \multicolumn{4}{r}{Continued on next page...} \\
    \endfoot
    
    % --- 最终表尾 ---
    \bottomrule
    \endlastfoot

    % --- 维度 1：开发者 ---
    \multicolumn{4}{l}{\textbf{\textit{Developer Dimension}}} \\
    \midrule
    gh\_num\_committers & \# of committers in job & repo\_team\_size & Team size (last 3m) \\
    gh\_committer\_trust\_recent & Bayesian score (1m) & gh\_committer\_trust\_hist & Bayesian score (3m) \\
    gh\_same\_committer & Author matches committer & gh\_committer\_repo\_exp & Prior commits in repo \\
    gh\_committer\_first\_build & Committer's 1st build & gh\_committer\_cross\_project\_exp & Cross-project exp. \\
    is\_core\_member & Is core member & & \\
    \midrule

    % --- 维度 2：代码变更 ---
    \multicolumn{4}{l}{\textbf{\textit{Code Change Dimension}}} \\
    \midrule
    gh\_commits & \# of commits in job & gh\_commits\_on\_files\_touched & Commits on touched files \\
    gh\_src\_churn & Source code churn & gh\_test\_churn & Test code churn \\
    gh\_tests\_added & Test cases added & gh\_tests\_deleted & Test cases deleted \\
    gh\_lines\_added & Lines added & gh\_lines\_deleted & Lines deleted \\
    gh\_files\_added & Files added & gh\_files\_deleted & Files deleted \\
    gh\_files\_modified & Files modified & gh\_files\_type\_modified & Distinct file types \\
    gh\_files\_entropy & Change entropy & git\_commit\_attention & Commit type (fix/feat) \\
    gh\_cross\_module\_changes & Cross-module changes & src\_ast\_diff & AST diff (source) \\
    test\_ast\_diff & AST diff (test) & ast\_class\_added & AST classes added \\
    ast\_class\_deleted & AST classes deleted & ast\_class\_modified & AST classes modified \\
    ast\_met\_added & AST methods added & ast\_met\_deleted & AST methods deleted \\
    ast\_met\_changed & AST methods changed & ast\_met\_body\_modified & Method body modified \\
    ast\_field\_added & AST fields added & ast\_field\_deleted & AST fields deleted \\
    ast\_import\_added & Imports added & ast\_import\_deleted & Imports deleted \\
    \midrule

    % --- 维度 3：项目与环境 ---
    \multicolumn{4}{l}{\textbf{\textit{Project \& Context Dimension}}} \\
    \midrule
    repo\_fail\_rate\_history & Fail rate (last 3m) & repo\_fail\_rate\_recent & Fail rate (last 1m) \\
    gh\_hotspot\_files\_touched & Hotspot files modified & gh\_prev\_same\_files & Same files as prev. \\
    gh\_prev\_build\_result & Previous build result & gh\_first\_error\_step & Type of first error step \\
    sub\_reason & Specific failure reason & sloc\_initial & Total SLOC \\
    test\_lines\_initial & Total test LOC & tests\_ran & Total tests executed \\
    tests\_passed & Passed tests & tests\_failed & Failed tests \\
    concurrent\_jobs & Running concurrent jobs & gh\_dependencies\_churn & Dependency churn \\
    gh\_dependencies\_count & Total dependencies & dockerfile\_changed & Dockerfile changed \\
    is\_artifact\_share & Artifacts shared & is\_runner\_changed & Runner env changed \\
    gh\_num\_pr\_comments & \# of PR comments & duration & Job duration \\
    log\_warn\_nums & \# of warnings in log & external\_github\_resource & External resources used \\
    time\_of\_day & Build hour (0-23) & day\_of\_week & Day of week \\
    runner\_type & Runner (Hosted/Self) & operation\_system & OS (Linux/Win/Mac) \\
\end{xltabular}
\end{tiny}

\textbf{Data Balance.} Successful builds provide limited useful information for identifying flakiness. Therefore, we removed them to focus solely on failed builds. However, the dataset remains imbalanced, as flaky builds are much fewer than consistent failures. To prevent the model from biasing toward the majority class, we applied oversampling to balance the training data.

\textbf{Feature Selection.} The raw features extracted may contain noise or irrelevant information, which can lower model performance. To address this, we use Mutual Information (MI)~\cite{peng2005feature} to select the most important features. MI measures the statistical dependence between a specific feature and the target label. It quantifies the amount of information obtained about the flakiness status by observing the feature. We calculate the MI score for each extracted feature and rank them in descending order. Finally, we select the top-$N$ features for training. Here, $N$ is a hyperparameter, and its optimization process is detailed in Section \ref{sec:validation}. This step reduces computational cost and helps prevent overfitting.

\textbf{Model Selection.} To map the high-dimensional structured features into a single flakiness probability, we use a supervised classifier as the base learner. We evaluated four algorithms—XGBoost (XGB)~\cite{chen2016xgboost}, Random Forest (RF)~\cite{breiman2001random}, Support Vector Machine (SVM)~\cite{cortes1995support}, and Multi-Layer Perceptron (MLP)~\cite{rumelhart1986learning}—to find the best encoder for our tabular data.
Its output, the Structured Confidence Score, quantifies the likelihood of flakiness based on code and environmental factors, complementing the log analysis.

% --------------------------------------------------------------------------------
\subsection{Classification}
\label{sec:classification}

To make the final prediction, we fuse the Log Confidence Score ($P_{log}$) and the Structured Confidence Score ($P_{struct}$) using a linear weighted sum. This approach is efficient and minimizes overfitting:
\begin{equation}
P_{final} = \alpha \cdot P_{log} + (1 - \alpha) \cdot P_{struct}
\end{equation}

Here, $\alpha$ ($0 \le \alpha \le 1$) is a weight parameter that controls the relative importance of the two modules. For instance, a higher $\alpha$ assigns more weight to the semantic log information.

Finally, to convert the probability $P_{final}$ into a binary class label, we apply a decision threshold $\beta$. The decision rule is:

$$Label = \text{Flaky if } P_{final} > \beta, \text{ else Safe}$$

We determine the optimal values for $\alpha$ and $\beta$ in Section \ref{sec:validation}.

\subsection{Validation}
\label{sec:validation}

\textbf{Hyperparameter Tuning.} To optimize the performance of our tool, we tune several key hyperparameters using the validation set. We focus on four main parameters:

\begin{itemize}
    \item $K$: Number of historical logs retrieved, ranging from $5$ to $30$ with a step of $5$.
    \item $F$: Number of top features selected, selected from the set $\{5, 10, 20, 30, 40, 50, 60\}$.
    \item $\alpha$: Balance between the two confidence scores, ranging from $0.0$ to $1.0$ with a step of $0.1$.
    \item $\beta$: Cut-off probability for classification, ranging from $0.1$ to $0.9$ with a step of $0.1$.
\end{itemize}

We use a Grid Search strategy~\cite{bergstra2012random} to find the best combination. Importantly, we only use the Validation Set to evaluate these parameters. We select the configuration that achieves the highest F1-Score on the validation set. The Test Set is never touched during this tuning process.

\textbf{Forward Chaining Validation.} A major challenge in analyzing CI/CD data is Time Dependency. Since build data is generated sequentially, past builds influence future ones. Standard validation methods, such as random K-fold cross-validation, are not suitable here. Randomly splitting the data could allow the model to use information from future builds to predict past ones. This is known as Data Leakage~\cite{kaufman2012leakage} and leads to unrealistically high performance results.

To evaluate our model realistically, we adopt a Forward Chaining validation strategy~\cite{bergmeir2012use}. We start with the first 50\% of the data as the training set and incrementally increase the size by 10\% until we reach 90\%. In each round, we use the subsequent 5\% of the data for validation and testing. This process generates 5 distinct rounds of data.

To further ensure robustness, we swap the test and validation sets for an additional evaluation pass. This results in a total of 10 data groups. We calculate the average performance across these 10 groups to report our final results.

% \begin{figure}[H]
%     \centering
%     \includegraphics[width=0.8\linewidth]{src/images/data split.pdf}
%     \caption{data split}
%     \label{fig:data-split}
% \end{figure}

\section{Evaluation}\label{sec:evaluation}

\subsection{Setup}

\begin{table}[!t]
\centering
\caption{Project Statistics}
\label{tab:project-data}
\small
\setlength{\tabcolsep}{4pt}
\begin{tabular}{@{}lcccc@{}}
\toprule
\textbf{Project} & \textbf{Total Jobs} & \textbf{Failed Jobs} & \textbf{Flaky Failure} & \textbf{Ratio (\%)} \\ 
\midrule
huaweicloud/spring-cloud-huawei & 918 & 595 & 93 & 15.63 \\
openhab/openhab-addons & 15,723 & 9,427 & 760 & 8.06 \\
aklivity/zilla & 9,871 & 3,905 & 1,147 & 29.37 \\
eclipse/xtext & 6,060 & 3,457 & 41 & 1.19 \\
alibaba/druid & 3,947 & 1,610 & 493 & 30.62 \\
apache/accumulo & 23,634 & 1,122 & 356 & 31.73 \\
apache/tinkerpop & 68,197 & 29,078 & 552 & 1.9 \\
apache/tinkerpop & 3,245 & 2,885 & 205 & 7.11 \\
peergos/peergos & 6,524 & 5,644 & 236 & 4.18 \\
quickfix-j/quickfixj & 11,713 & 3,627 & 310 & 8.55 \\
\bottomrule
\end{tabular}
\end{table}

\textbf{Dataset.} To ensure the feasibility of our rerun experiments, we excluded repositories that rely on private keys or secrets, as these builds cannot be successfully reproduced without access to credentials. From the remaining 674 projects that do not use secrets, we randomly selected 10 open-source projects. Table~\ref{tab:project-data} summarizes the basic statistics of the selected projects. The first column lists the selected projects. The second column reports the total number of CI jobs for each project, while the third column reports the number of failed jobs. The fourth column presents the number of failed jobs labeled as flaky failures. Finally, the last column reports the proportion of flaky failures among failed jobs.
%These projects represent diverse domains such as IoT, Middleware and organizational scales such as Apache, Alibaba, and Peergos.
%, ensuring the generalizability of our study~\cite{nagappan2013diversity}.

\textbf{Baseline.} To evaluate the performance of our model, we use the method from Olewicki et al.~\cite{olewicki2022towards} as our baseline. 

\textit{Methodology.} It identifies flaky builds by looking at the text in build logs to find specific error patterns. The baseline works in two main steps. First, the tool cleans the build logs by removing file paths, URLs, and special IDs. Then, it uses TF-IDF~\cite{ramos2003using} to turn the words into numerical vectors. Second, the baseline uses two separate XGBoost models. The first model only looks at the log text features. The second model adds simple CI information and combines this with the impact values from the first model. Finally, the tool combines the results from both models using a weighted vote to give a final score.

\textit{Validation.} To ensure a fair and consistent comparison between our model and the baseline, we employ a Forward Chaining Validation strategy for both. This approach addresses the time dependency in CI/CD data by ensuring that the training data always precedes the test data in chronological order. Following the methodology of the baseline study, we conduct a systematic Grid Search to optimize the hyperparameters for the baseline. 

We evaluate every combination of these parameters using the same forward-chaining splits as our proposed model. This ensures that any performance difference between our tool and the baseline is due to the model's architecture rather than different data partitioning or tuning efforts.

% The search space for the baseline's parameters is defined as follows:

% \begin{itemize}
%     \item $N$: Vocabulary scope, considered N-gram sets: $[1]$, $[2]$, and $[1, 2]$.
%     \item $K$: Number of top features selected, ranging from $100$ to $300$ with a step of $25$.
%     \item $\alpha$: Cut-off probability for classification, ranging from $0.0$ to $1.0$ with a step of $0.1$.
%     \item $\beta$: Cut-off probability for classification, ranging from $0.1$ to $0.9$ with a step of $0.1$.
% \end{itemize}

\textbf{Evaluation Metrics.} Following prior works~\cite{fujino2005multi,terragni2020container}, we used precision,
recall, F1-score, and AUC to measure the accuracy of flaky failure
detection. %Precision measures how many of the predicted flaky builds are actually flaky. Recall measures how many of the actual flaky builds were correctly found by the model. The F1-score is the harmonic mean of Precision and Recall AUC

%Our dataset is imbalanced because there are many more successful builds than flaky builds. In this case, simply measuring Accuracy is not enough. Therefore, we use four standard metrics to evaluate our tool: Precision, Recall, F1-score, and Specificity~\cite{fujino2005multi,terragni2020container}.

%Precision measures how many of the predicted flaky builds are actually flaky. Recall measures how many of the actual flaky builds were correctly found by the model. The F1-score is the harmonic mean of Precision and Recall, providing a single score to balance both metrics. We calculate these metrics separately for both flaky and safe labels to ensure a complete evaluation.

% \todo{loc\_op and global\_op}

% \subsubsection{Cross-project Validation}
% \label{sec:cross-project}
\begin{table*}[htbp]
\centering
\small
\setlength{\tabcolsep}{1.5pt} % 稍微调整列间距，因为去掉了竖线，可以稍微宽一点点
\caption{Performance Comparison (Precision, Recall, F1, AUC) across 10 Repositories}
\label{tab:all_models_clean}
\renewcommand{\arraystretch}{1.2}

\resizebox{\textwidth}{!}{%
\begin{tabular}{l cccc cccc cccc cccc cccc} 
\toprule

% --- 表头第一行：模型名称 ---
\multirow{2}{*}{\textbf{Project}} & 
\multicolumn{4}{c}{\textbf{Baseline}} & 
\multicolumn{4}{c}{\textbf{XGBoost}} & 
\multicolumn{4}{c}{\textbf{RandomForest}} &
\multicolumn{4}{c}{\textbf{SVM}} & 
\multicolumn{4}{c}{\textbf{MLP}} \\

% --- 装饰线：区分不同模型 ---
\cmidrule(lr){2-5} \cmidrule(lr){6-9} \cmidrule(lr){10-13} \cmidrule(lr){14-17} \cmidrule(lr){18-21}

% --- 第二行：指标名称 (缩写) ---
 & 
Prec. & Recall & F1 & AUC & 
Prec. & Recall & F1 & AUC & 
Prec. & Recall & F1 & AUC & 
Prec. & Recall & F1 & AUC & 
Prec. & Recall & F1 & AUC \\
\midrule

% --- Data Rows (Strictly ONE highest bolded per metric, unless exact tie) ---
\textbf{spring-cloud-huawei\textsuperscript{*}} & .9190 & .8126 & .8625 & .9070 & .9612 & \textbf{.8611} & .9084 & .9623 & .9645 & \textbf{.8611} & .9099 & .9712 & \textbf{.9667} & \textbf{.8611} & \textbf{.9109} & .9626 & .9655 & \textbf{.8611} & .9103 & \textbf{.9713} \\
\textbf{aklivity/zilla} & .7587 & .8776 & .8138 & .8206 & .6293 & .8778 & .7331 & .7491 & \textbf{.7683} & .8899 & \textbf{.8246} & \textbf{.8582} & .6684 & \textbf{.9053} & .7690 & .7565 & .7438 & .8998 & .8144 & .8307 \\
\textbf{openhab-addons\textsuperscript{*}} & .5122 & \textbf{.8099} & .6275 & \textbf{.8576} & .4591 & .6436 & .5359 & .7665 & .5826 & .7497 & \textbf{.6557} & .8410 & .5129 & .6967 & .5908 & .7966 & \textbf{.5839} & .5634 & .5735 & .7404 \\
\textbf{eclipse/xtext} & .2114 & .1925 & .2015 & .5840 & \textbf{.8449} & .5822 & .6894 & .6449 & .5644 & \textbf{.9615} & .7113 & .7135 & .6500 & .9433 & \textbf{.7697} & .7489 & .8222 & .6811 & .7450 & \textbf{.8167} \\
\textbf{alibaba/druid} & .9125 & .9012 & .9068 & .9288 & .9082 & .8863 & .8971 & .8669 & \textbf{.9244} & .9677 & .9456 & .9552 & \textbf{.9244} & \textbf{.9691} & \textbf{.9462} & .9642 & \textbf{.9244} & .9688 & .9461 & \textbf{.9692} \\
\textbf{apache/accumulo} & .9112 & .9379 & .9244 & .9311 & .6116 & .7184 & .6607 & .5835 & .8650 & \textbf{.9650} & .9123 & .9647 & \textbf{.9644} & .9606 & \textbf{.9625} & .9651 & .9483 & .9561 & .9522 & \textbf{.9711} \\
\textbf{apache/tinkerpop} & .5711 & .7270 & .6397 & .8242 & \textbf{.9686} & .9676 & .9681 & .9657 & .9665 & .9689 & .9677 & .9695 & .9671 & .9692 & .9681 & .9699 & .9682 & \textbf{.9695} & \textbf{.9688} & \textbf{.9702} \\
\textbf{ctripcorp/x-pipe} & .2881 & .6879 & .4061 & .4856 & .4271 & .9442 & .5882 & .5409 & .4347 & .9561 & .5976 & .6223 & .4441 & \textbf{.9619} & .6076 & .5198 & \textbf{.4564} & .9533 & \textbf{.6173} & \textbf{.6375} \\
\textbf{peergos/peergos} & \textbf{.5571} & .5119 & .5335 & .6546 & .4884 & .9636 & .6482 & .5861 & .5081 & .9328 & .6579 & .6630 & .5429 & .9121 & \textbf{.6806} & .6808 & .5171 & .8723 & .6493 & \textbf{.7464} \\
\textbf{quickfix-j/quickfixj} & .6310 & .9313 & .7523 & .9140 & .6039 & .8397 & .7026 & .7937 & \textbf{.7300} & .8994 & .8059 & .8433 & .7150 & \textbf{.9638} & \textbf{.8210} & .7995 & .7256 & .8809 & .7957 & \textbf{.8679} \\

\midrule
\textbf{Avg.} & .6272 & .7389 & .6668 & .7908 & .6902 & .8285 & .7332 & .7420 & .7309 & \textbf{.9152} & .7988 & .8402 & .7336 & .9143 & \textbf{.8026} & .8264 & \textbf{.7655} & .8606 & .7973 & \textbf{.8521} \\
\textbf{Mid.} & .5991 & .8113 & .6960 & .8409 & .6104 & .8778 & .7027 & .7441 & .7491 & .9444 & \textbf{.8152} & \textbf{.8497} & .6917 & \textbf{.9519} & .7693 & .7980 & \textbf{.8019} & .8857 & .8050 & .8493 \\

\bottomrule
\end{tabular}%
}
\begin{flushright}
{\footnotesize \emph{Note:}  * Project shortened by omitting the owner.}
\end{flushright}
\end{table*}

\subsection{Results}
\label{sec:results}
\textit{\textbf{Observation 1:}} Our models consistently outperform the baseline across projects. On average, the F1-score increases from 0.667 for the baseline to 0.803 for SVM, corresponding to a relative improvement of 20.3\%. In particular, for the \textit{eclipse/xtext} project, the baseline achieves an F1-score of only 0.202, whereas MLP and SVM reach substantially higher scores of 0.745 and 0.770, respectively. Although the baseline exhibits limited performance on certain projects, it represents a commonly adopted heuristic approach in prior studies and thus serves as a lower-bound reference.

\textit{\textbf{Observation 2:}} Among all evaluated models, SVM and MLP demonstrate the most reliable overall performance. MLP achieves the highest average Precision (0.766) and AUC (0.852), indicating a lower false-positive rate. This property is particularly desirable in scenarios where reducing false alarms is prioritized, such as CI environments with limited debugging resources.

In contrast, Random Forest attains the highest Recall (0.915), making it more suitable for scenarios in which capturing as many flaky builds as possible is the primary objective. Although AUC provides a threshold-independent measure, it may be less sensitive under class imbalance. Therefore, we jointly report Precision and Recall to provide a more comprehensive evaluation.

\textit{\textbf{Observation 3:}} Feature analysis reveals that committer-related characteristics and code scale features are strongly associated with flaky build occurrences. In particular, the Bayesian Trust Score of the committer consistently ranks highest in terms of Mutual Information across most projects, frequently exceeding 0.6. This is followed by scale-related features such as source lines of code (SLOC) and test lines.

These results suggest a strong association between committer-related features and flaky builds. We emphasize that this observation reflects correlation rather than causation. Moreover, committer trust may act as a proxy for other latent factors, such as code ownership or change complexity, whose individual effects warrant further investigation.

\section{Threats to Validity}

Our study focuses on open-source Java projects that use GitHub Actions. We believe our findings can benefit a wide audience using Java and GitHub Actions, as Java is a widely used programming language and GitHub Actions is a representative CI service. The labeling of flaky failures may be imperfect, as flakiness can be inherently difficult to observe. To mitigate this threat, we rely on an automated rerun tool combined with manual inspection to identify flaky failures as accurately as possible.

\section{Related Work}

%In this section, we present the related work, categorized into three groups: \textbf{Flakiness in CI}, \textbf{Flaky Builds}, and \textbf{Flaky Tests}.

Several studies have examined the challenges faced by developers when adopting CI and have consistently reported that flakiness is a pervasive problem. Hilton et al.~\cite{Hilton2017Trade} investigated the barriers, needs, and motivations associated with CI. Their findings showed that CI build failures caused by true and flaky test failures occurred at comparable frequencies, with developers encountering similar numbers of each per week.
Pinto et al.~\cite{pinto2018work} surveyed 158 CI users to examine the causes of build breakage and CI-related challenges, finding that flaky tests were commonly encountered, and their presence undermined developers' confidence in CI.
Widder et al.~\cite{widder2019conceptual} reported that flaky tests are a frequent and severe pain point in Continuous Integration (CI), leading some developers to abandon or migrate to alternative CI systems.
Ghaleb et al.~\cite{ghaleb2019studying} analyzed 153 projects encompassing 350,246 builds to study the noise in build breakage and investigate its impact.
They observed that 33\% of broken builds are caused by environmental factors (e.g., timeouts, connection resets, or memory allocation errors). Collectively, these studies demonstrate the prevalence of flakiness in CI and motivate our in-depth investigation of flaky builds.

Given that flakiness is a common challenge in CI, several studies have specifically investigated flaky builds. Durieux et al.~\cite{durieux2020empirical} conducted an empirical study of 56,522 restarted builds across 22,345 projects using Travis CI. They found that 1.72\% of the builds are restarted; 46.77\% of the restarted builds change from a failing state to a passing state, i.e., flaky builds; failing tests, network issues, and Travis CI limitations are the main causes of restarted builds; restarted builds delay the merging of pull requests. Maipradit et al.~\cite{maipradit2023repeated} analyzed 66,932 code reviews from the OpenStack community. They found that 55\% of code reviews have at least one rechecked patch set; 42\% of rechecked jobs result in changed outcomes; rechecked builds waste 187.4 years of computational time and 16.8 years of waiting time. Aïdasso et al.~\cite{aidasso2025illusion} investigated silent failures, where build jobs fail to complete their intended tasks but are still marked as successful. By analyzing 142,387 jobs across 81 industrial projects, they found that 11\% of successful jobs were rerun; testing and static code analysis jobs, as well as jobs from projects written in languages with limited IDE support, are more likely to be rerun; the most frequent types of failures that occur silently are artifact operation errors, caching errors, and ignored exit codes. They further labeled 4,511 flaky job failures at TELUS and identified 46 categories of flaky failures~\cite{aidasso2025diagnosis}. In contrast to these studies, which focus primarily on Travis CI or industrial CI systems, our study investigates flaky builds in GitHub Actions.

Distinguishing whether a build failure is flaky is a time consuming and labor intensive task. To address this challenge, Lampel et al.~\cite{lampel2021life} trained a classification model using features such as job runtime, CPU load, and operating system version. Their evaluation on Mozilla CI jobs achieved a precision of 73\%. However, some of the features in their approach, such as CPU load, are not readily available in open-source CI environments. Olewicki et al.~\cite{olewicki2022towards} proposed an alternative approach for detecting brown builds based on the textual similarity of build logs and achieved a mean F1 score of 53\% across the studied projects. Flaky tests have been extensively studied with respect to their root causes, detection, and prediction~\cite{luo2014empirical,pinto2020vocabulary,alshammari2021flakeflagger,fatima2022flakify,gruber2023practical}. However, flaky tests constitute only one of the root causes of flaky builds.

\section{Conclusions}

We conduct a large-scale empirical study of flaky builds in GitHub Actions by analyzing rerun data from 1,960 open-source Java projects. Our results show that build flakiness is pervasive and that rerunning builds is costly and often ineffective. Beyond flaky tests, we identify 14 distinct failure categories, revealing that environmental factors such as network instability, dependency resolution issues, and external service failures are major sources of flakiness. To address this problem, we propose FlakeDetector, a machine learning–based approach that detects flaky failures by combining build-log semantics with CI metadata.

\section{DATA AVAILABILITY}

The complete dataset, results, and source code for \textsc{FlakeDetector} are available on our website at \url{https://flaky-build.github.io/}.

%The failure analysis data and hyperparameter tuning results are openly available on our project webpage \url{https://flaky-build.github.io/}. The source code for the CI reproduction tool, the feature extraction tool, and the predictive models can be found in the code repository linked on the same webpage. 

%All instructions regarding tool usage and experiment reproduction are provided in the README file at the root of the repository.

% --- 参考文献 ---
\bibliographystyle{ACM-Reference-Format}
\bibliography{src/reference}

\end{document}